\documentclass[conference]{IEEEtran}
\usepackage{cite}
\ifCLASSINFOpdf
  \usepackage[pdftex]{graphicx}
  \graphicspath{{./}}
\else
\fi

\begin{document}
\title{A comparison of C-shaped and brush armature performance}


\author{\IEEEauthorblockN{Barbara Wild\IEEEauthorrefmark{1}, Farid Alouahabi\IEEEauthorrefmark{1}, Dejan Simicic\IEEEauthorrefmark{1}, Markus Schneider\IEEEauthorrefmark{1} and Ryan Hoffman\IEEEauthorrefmark{2}}
\IEEEauthorblockA{\IEEEauthorrefmark{1}French-German Research Institute of Saint Louis, France}
\IEEEauthorblockA{\IEEEauthorrefmark{2}Office of Naval Research, Arlington, VA 22203 USA}
}


%Telephone: (800) 555--1212, Fax: (888) 555--1212}


\maketitle

\begin{abstract}

The most important part of a  railgun launch package is the armature where the  electromagnetic force is generated leading to the acceleration of the launch package.   In case of metal armatures, the most commonly used armature types are the C-shape and the multi-fiber brush technology. However, rarely both armature types were systematically compared under similar experimental conditions. That is why we constructed  launch packages   based on the C-shaped and brush armature technology with comparable armature and payload mass. With these launch packages a series of experiments were performed in an energy range between 0.8 MJ and 1.13 MJ corresponding to a speed range between 950 m/s and 1400 m/s. The results of the experiments were then analyzed qualitatively and quantitatively. On the one hand our results show that the total losses are higher for the C-shaped armature technology than for the brush aramture technology. On the other hand our results show that launch packages based on the C-shaped technology convert better electrical energy into kinetic energy.  

\end{abstract}
% \fi
%\IEEEpeerreviewmaketitle
\section{Introduction}
Launch packages of electromagnetic launchers consist of the projectile and/or the sabot and the armature. The armature is the part where the electromagnetic force is generated and therefore the most essential part of the launch package: it carries the full accelerating current and is subject to the full accelerating forces. If the armature fails, i.e. if the electrical contact between the armature and the rails is lost, plasma is developed which might lead to the erosion of the rails. There are different options of possible armature designs, but the most common and most studied designs are the so called C-shaped armatures and brush armatures \cite{Barber2003,Schneider2003a,Schneider2005,Schneider2005a}. But so far, to our knowledge,  both types of armature were not studied under the same experimental conditions, in order to determine which  performs better.  That is why, we fabricated launch packages based on the  C-shaped and brush armature technology which were similar in terms of armature and payload mass. With these launch packages  we performed single shot experiments with the {RA}pid {FI}re {RA}ilgun (RAFIRA) in an energy range between 0.8 MJ and 1.13 MJ corresponding to a speed range between 950 m/s and 1400 m/s. The contact behavior during the shot was studied using sophisticated metrology: muzzle voltage probe, Doppler radar system, and X-ray flash radiography. The results obtained from our experiments enabled us to analyze qualitatively  and quantitatively the losses and the efficiencies of both types of armatures.  

\section{Experimental set-up}
 \subsection{Railgun RAFIRA}
The experiments described in this paper were performed with  the launcher {RAFIRA} (Rapid Fire Railgun)\cite{Schneider2009a,Schneider2009}. This linear electromagnetic accelerator is equipped with rectangular rails having a caliber of 25~$\times$~25 mm$^2$ and a length of about 3 m. Due to its open-bore structure consisting of glass-fiber reinforced plastic (GRP) bars, the rails are easily  accessible and  the metrology used for the characterization of the shots can be implemented without any major problems. Up to 3.06\,MJ in total can be provided to {RAFIRA} as primary energy. More technical details and images of the equipment used in the experiments can be found in \cite{Schneider2009a,Schneider2009}. The  metrology which is 
used in the experiments described below includes a muzzle voltage probe, a radar system for projectile velocity determination at the muzzle and X-ray flash radiography.
\begin{figure}
  \centering
    \includegraphics[scale= 0.5]{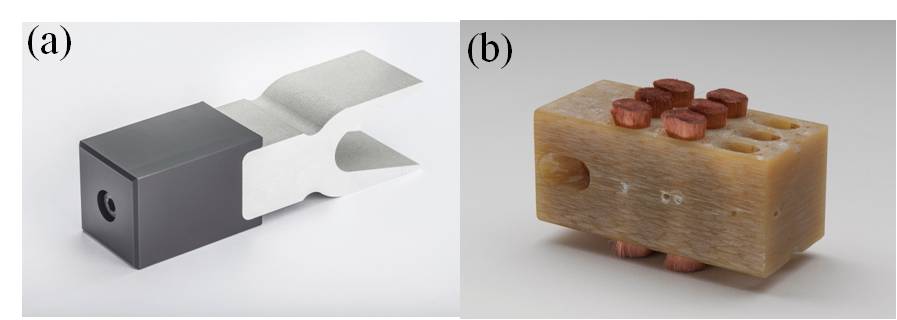}
    \caption{Typical projectiles used in the experiments: (a) C-shaped projectile, (b) brush projectile}
  \label{fig_projectile}
\end{figure}
 \subsection{Projectiles}
For our experiments two types of launch packages characterized by different armatures were used. But for reasons of simplicity only the term projectile will be used for the whole launch package. Both projectiles were constructed in such a way to be as similar as possible in terms of armature and payload mass for a given action integral.
\subsubsection{C-shaped projectile}
	The C-shaped projectile  is made of aluminum and is adapted to the caliber of {RAFIRA} (s. Figure \ref{fig_projectile}(a)). The mass of the C-shaped armature is about $m_a~=~30~g$. The bore rider (black block, figure \ref{fig_projectile}(a)) is made of Lexan  with a mass of about  $m_p~=~50~g$ 
leading to a total mass of the whole projectile of about 80~g.
\subsubsection{Brush projectile}
This type of projectile consists of  six  brushes made of many copper fibers ($m_a \approx$ 40 g) incorporated into a sabot ($m_p \approx$ 40 g) made of GRP serving as armature (see figure \ref{fig_projectile}(b))\cite{Schneider2003a,Schneider2005a,Schneider2005} leading to a total weight of 80~g as well. 

Please note that the difference in weight of the C-shaped and brush armatures is due to material properties. According to \cite{Barber2003, Marshall2004a}, the minimum armature mass needed for solid railgun armatures is limited by resistive heating of the metal. This resistive heating is best characterized by the action integral $\int I^2 dt$ and the action constant $g_1$, a material property (which is for copper $g_1 = 80.5~kA^2s/mm^4$ and for aluminum  $g_1 = 19~kA^2s/mm^4$) \cite{Barber2003, Marshall2004a}. The two are related by:
\begin{equation}
\label{equ_action}
\int I^2 dt = g_1 A^2
\end{equation}
where $A$ is the current carrying cross-section. Using equation (\ref{equ_action}), we obtain for a minimum armature mass: 
\begin{equation}
\label{equ_ma}
m_a = \rho b A = \rho b \sqrt{\frac{\int I^2 dt}{g_1}}
\end{equation}
where $b$ is the railgun bore and $\rho$ the density. For a given action integral  and railgun bore the minimum mass armature depends on $\rho /\sqrt{g_1}$. That is why, for the same  action integral, the mass armature of copper armatures is always higher than the mass of aluminum armatures ($\rho_{Cu} /\sqrt{g_1} = 0.032~kg/As^{\frac{1}{2}},~ \rho_{Al}/\sqrt{g_1} = 0.017~kg/As^{\frac{1}{2}}$ \cite{Marshall2004a}).

\subsection{Experiments}
\begin{figure}
  \centering
    \includegraphics[width=2.5in]{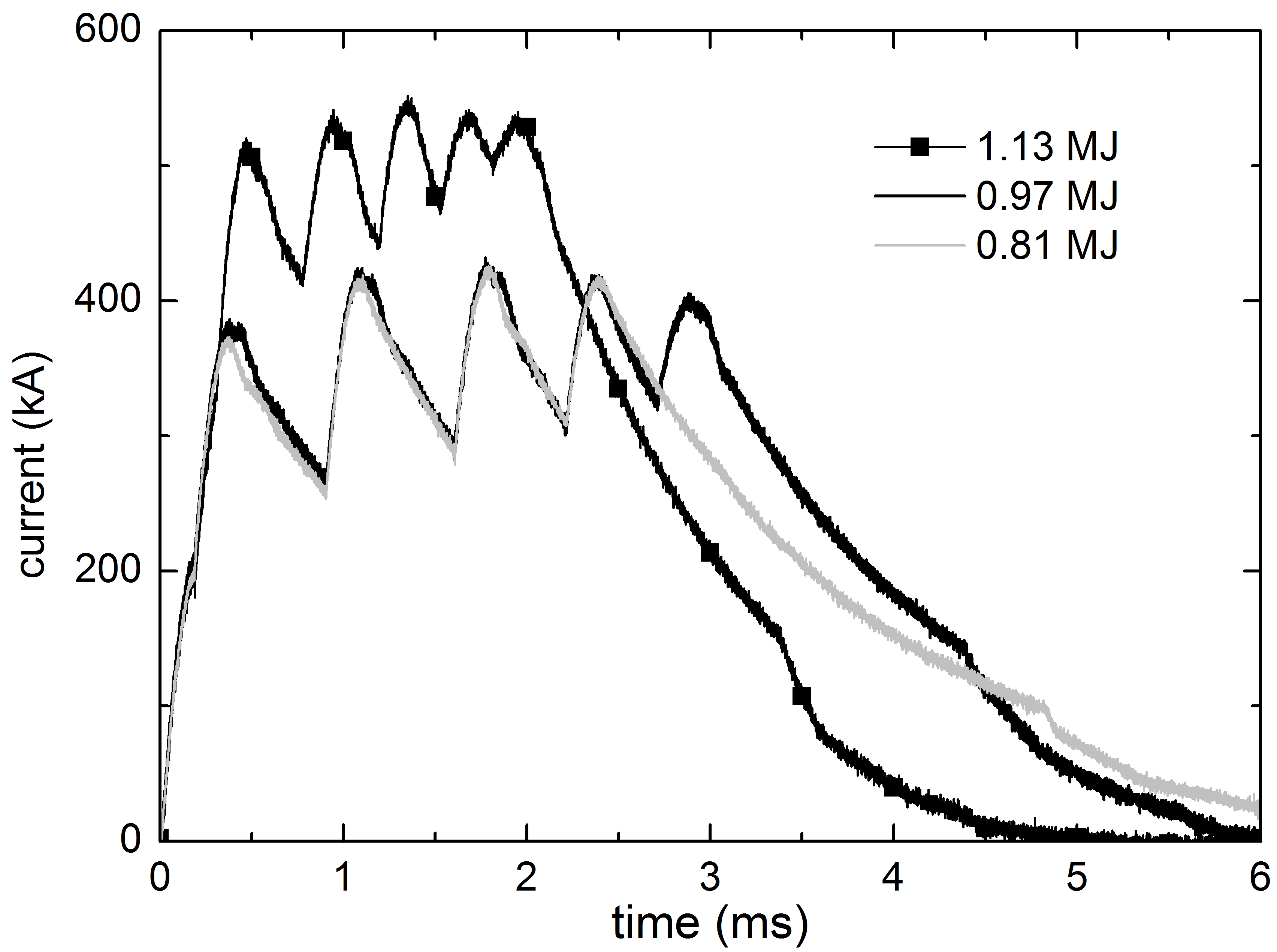}
    \caption{Profiles of current pulses used for the acceleration of both types of projectiles for  applied primary energies of 0.81 MJ, 0.97 MJ and 1.13 MJ}
  \label{fig_pulse}
\end{figure}

In total we performed a series of 6 single shots.  For these shots the applied primary energy was varied from 0.81\,MJ to 1.13\,MJ in steps of 162\,kJ.  The applied current pulses used for both types of projectiles are shown in figure \ref{fig_pulse}. Before each shot either new copper (CuCr) rails (for experiments with C-shaped projectiles) or new aluminum (Dural) rails (for experiments with brush projectiles) were mounted into the  launcher. In this way  the same experimental conditions for each experiment and a good electrical contact between the corresponding projectiles and the rails  were assured \cite{Wild2015a}. 
\section{Experimental results and discussion}
\subsection{Muzzle voltage}
\label{sub_muzzle}
\begin{figure}
  \centering
    \includegraphics[width=\columnwidth]{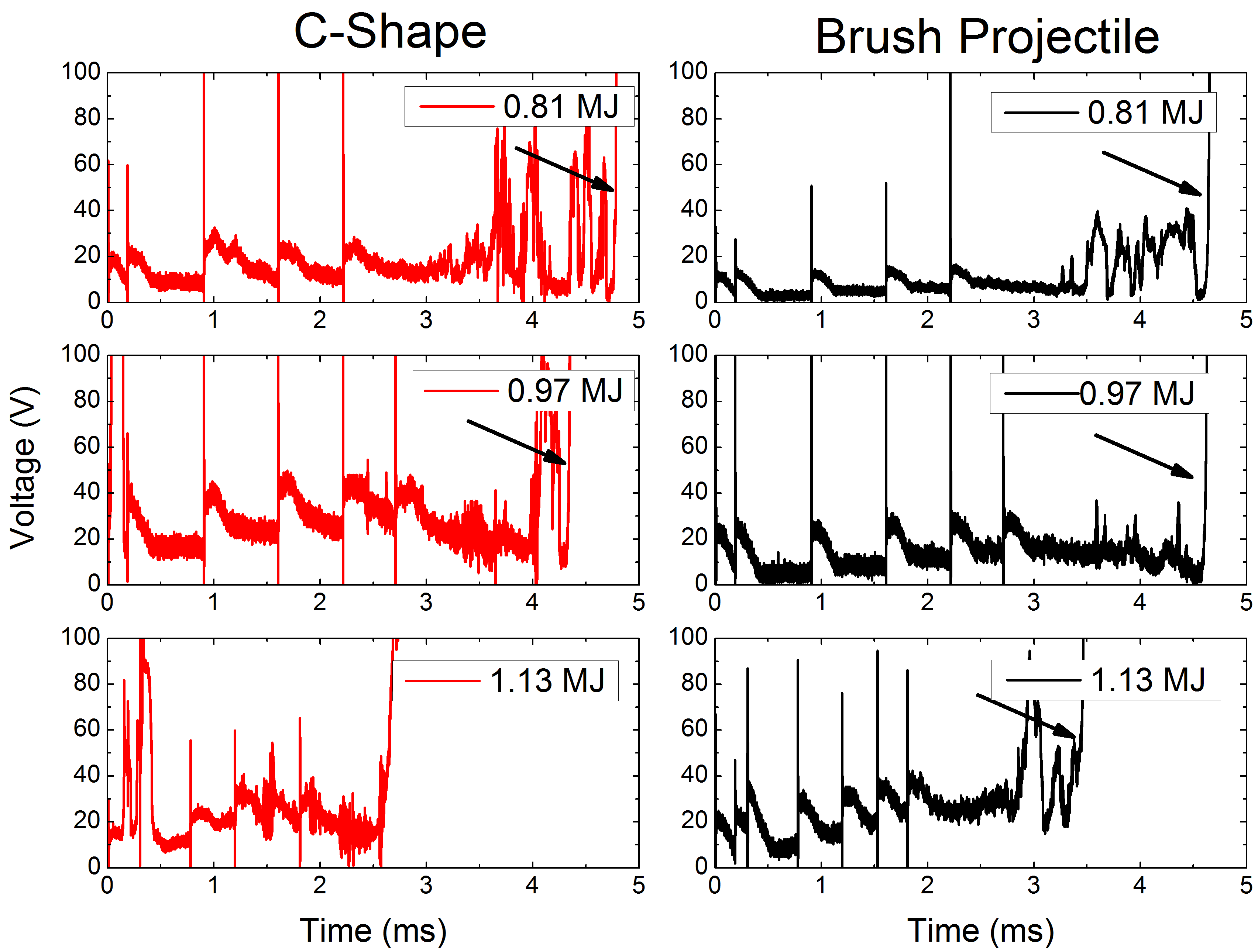}
    \caption{Muzzle voltage profiles of each shot measured for C-shaped projectiles (left) and brush projectiles (right) and increasing primary energy}
  \label{fig_muzzle}
\end{figure}

The measured muzzle voltage profiles of all shots are shown in figure \ref{fig_muzzle}. The narrow peaks at the beginning of the profiles are due to the discharge of the capacitor units. The number of these peaks corresponds to the number of the discharged capacitor units. The large peak at the end of each voltage profile (U\,$>$\,50 V, see arrows in figure \ref{fig_muzzle})  corresponds to the muzzle arc which is formed when the projectiles exits the railgun. In general,  a good sliding contact is characterized by voltage amplitudes below 50\,V, whereas voltage amplitudes over this value indicate a loss of the solid contact between rails and armature.  
In case that the solid contact loss leads to an electric arc, the term contact transition is used. 
\subsubsection{Brush projectiles}
At the lowest primary energy (E\,=\,0.81\,MJ) the average voltage amplitude U$_M$ from t\,=\,0\,ms until t\,=\,3.5\,ms is about 4\,V, indicating a good contact behavior. At t = 3.5 ms peaks due to current switching between the different brush armatures appear \cite{Schneider2003a}. These peaks disappear at t \,= \,4.5\,ms and  the solid contact is re-established before the exit of the projectile at t\,=\,4.7\,ms. At E\,=\,0.97 MJ, the voltage profile shows an overall  good contact  behavior (U$_M\,\approx$\,12\,V) with only minor peaks due to  current switching process between brushes at around t\,=\,3.6\,ms. The projectile leaves  the launcher at t\,=\,4.6\,ms. At the highest primary energy, E\,=\,1.13\,MJ,  U$_M$ is about 20\,V between t\,=\,0\,ms and t\,=\,3\,ms. At t\,=\,3\,ms  peaks due to the current switching between the different brush armatures start to arise, but no contact transition is observed before the exit of the projectile  at t\,=\,3.4\,ms. 
\subsubsection{C-shaped projectile}
\begin{figure}
  \centering
    \includegraphics[scale=0.4]{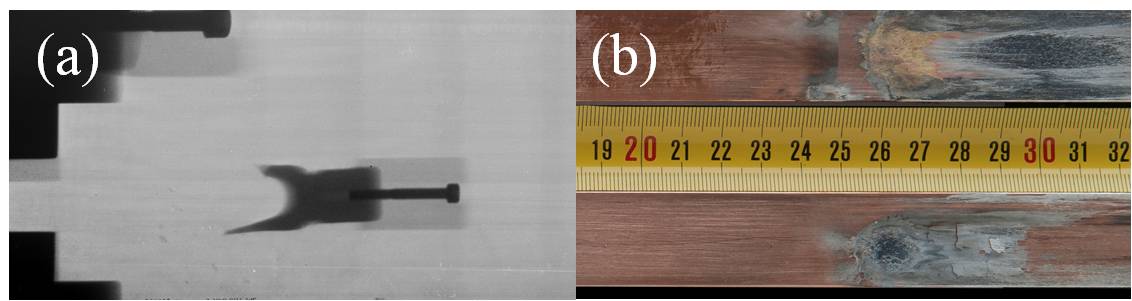}
    \caption{(a) X-ray photograph of the C-shaped projectile leaving the launcher RAFIRA, (b) part of the post shot CuCr rails, the shown part is the starting position of the C-shaped projectile, applied energy, 0.97 MJ}
  \label{fig_rx_rail}
\end{figure}

At the lowest primary energy (E\,=\,0.81\,MJ) U$_M$ is about 10~V between t\,=\,0\,ms and t\,=\,3.5\,ms. At t\,=\,3.5\,ms  narrow peaks with amplitudes of up to 80 V appear indicating a temporally loss of the solid contact which is re-established before the exit of the projectile at t\,=\,4.8\,ms. These peaks might be attributed to a current switching between local contact zones on the C-shaped armature/rail interface due to material loss and/or the decrease of the in-bore magnetic field, leading to a drop of the compressive force on the armature legs \cite{Barber2003}. 
At E\,=\,0.97\,MJ,  between t\,=\,0\,ms and t\,=\,4.0 \,ms U$_M$ is about 10~V to 30\,V indicating a good contact behavior. At 4.0 ms a peak  with a width of $\Delta$t \,=\,0.25 ms and a amplitude of U$_M>$\,50 V indicates a temporally contact loss. The solid contact is re-established before the exit of the projectile  at t~=~4.3\,ms. Additionally to this peak a peak at t\,$\approx$\,0.1 ms appears, which is  in correlation with the post-shot state of the rail surfaces (see figure \ref{fig_rx_rail}(b)) showing traces of molten aluminum on the rails at the start position. Therefore this peak is due to melting of armature material at the beginning of the shot.
At the highest primary energy, E\,=\,1.13\,MJ, the average muzzle voltage U$_M$ is between 10\,V and 30~V. At t\,=\,2.6\,ms the C-shaped armature undergoes a contact transition, because one of the trailing arms breaks during the experiment (see figure \ref{fig_rx_rail}(a)). Similar to the primary energy of 0.97\,MJ, a peak at around t\,=\,0.2\,ms appears as well because of the melting of armature material at the beginning of the shot. 

Please note, that U$_M$ is always higher for the C-shaped projectile than for the brush projectile. This can be explained by a very simple model:   In the case of a C-shaped armature one can assume that the current flows through an area of $A$\,=\,3.4\,mm\,$\times$\,25~mm~=~85~mm$^2$, (3.4\,mm is the approximately skin depth of the current and 25 mm is the width of the C-shaped armature) and a length of $l$\,=\,85\,mm (corresponding roughly to the circumference of the C-shaped armature). Knowing the electrical resistivity of  aluminum ($\rho_{el}^{Al}=2.8\,\times\,10^{-8}\,\Omega m$) the electrical resistance of the C-shaped armature can be calculated by:
\begin{equation}
R = \rho \frac{l}{A}
\label{equ_r}
 \end{equation}
Now using Ohm's law with I = 400 kA
\begin{equation}
U_M = R \times I
\label{equ_ohm}
\end{equation}
a muzzle voltage of U$_M \approx$ 10 V is obtained.
In the case of  brush armatures  $A$ is assumed to be 80\,mm$^2$ (corresponding to an area of three brushes with a diameter of 5.8\,mm) and  $l$ = 25 mm (corresponding to the caliber size). Using equations (\ref{equ_r}) and (\ref{equ_ohm}) and  $\rho_{el}^{Cu}\,=\,1.7\,\times\,10^{-8}\,\Omega\,m$ one obtaines U$_M \approx$ 2V. 
In the case of the C-shaped armature the higher muzzle voltage is first of all due to the longer current path in the C-shaped armature ($l_{Cshape}$ = 3.4 $\times l_{brush}$) and second due to the higher electrical resistivity of aluminum  ($\rho_{el}^{Al}$ = 1.65 $\times$ $\rho_{el}^{Cu}$).

\subsection{Velocities}
 Because of the installed Doppler radar system at the muzzle of RAFIRA, the velocity of both types of projectiles for each shot were determined with a very high accuracy\cite{Schneider2003}. 
The results of all experiments are summarized in table \ref{tab_velo}.
 The muzzle velocity increases with applied energy from 955 m/s  of up to 1435 m/s for the C-shaped projectile, whereas for the brush projectile the muzzle velocity increases from 1027~m/s up to 1381~m/s. Except for the first experiment the velocity of the C-shaped projectile is always greater ($\Delta v_{max} = 70~m/s$) than the muzzle velocity of the brush projectile.  
    \begin{table}
		\centering
        \caption{Experimental and theoretical velocities}
        \label{tab_velo}
        \begin{tabular} {| c | c c |c c| }
            \cline{1-5}
						 \cline{1-5}
           & \multicolumn{2}{|c|}{\textbf{$v_{exp}$ (m/s)}}& \multicolumn{2}{|c|}{\textbf{$v_{theo}$ (m/s)}}\\
            \cline{1-5}
						 %\cline{1-3}
              \textbf{Energy (MJ)} & C-shape & brush & C-shape & brush\\
						\hline
						%   & \multicolumn{2}{|c|}{Muzzle velocities (m/s)} \\
            %\hline
          	0.81 &	955 & 1027 & 1083& 1075\\
			\hline%
				0.97 & 1214 & 1120 & 1207 & 1218\\
			\hline%
				1.13 & 1435 & 1381 & 1540& 1501\\
            \hline
						 \cline{1-5}
        \end{tabular}
    \end{table}	
\subsection{Post shot rail surfaces}
The state of the post shot-state of the rail surfaces is an excellent source of information about the sliding contact behavior, because very often state of the post surfaces can be correlated to contact transitions.

\begin{figure}
  \centering
    \includegraphics[scale=0.5]{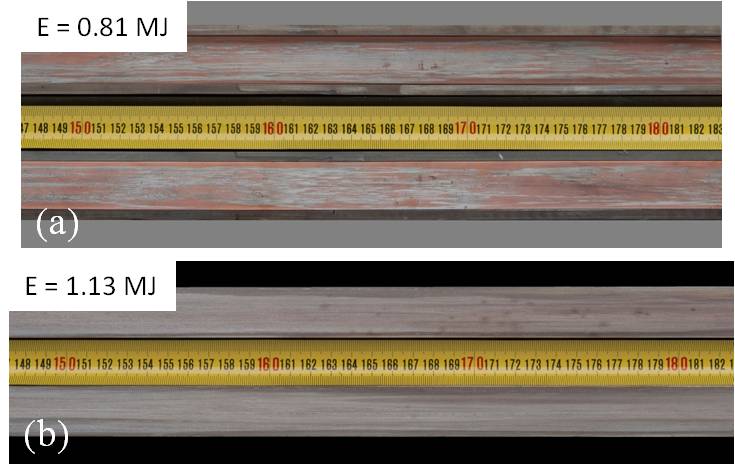}
		\includegraphics[scale=0.5]{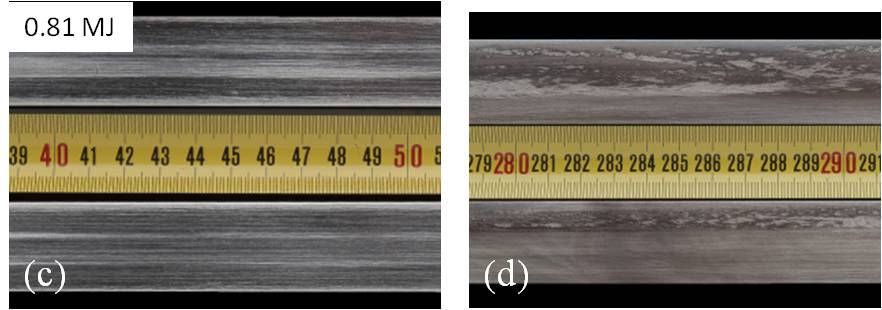}
    \caption{Part of CuCr and Dural rail surfaces after experiments;  Copper rail, C-shaped projectile and applied energy:(a) 0.81 MJ and (b) 1.13 MJ kJ, Dural rails, applied energy 0.81 MJ (c) at the beginning of the rails (d) at the end of the rails}
  \label{fig_rails}
\end{figure}

\begin{figure}[t]
  \centering
    \includegraphics[scale=0.3]{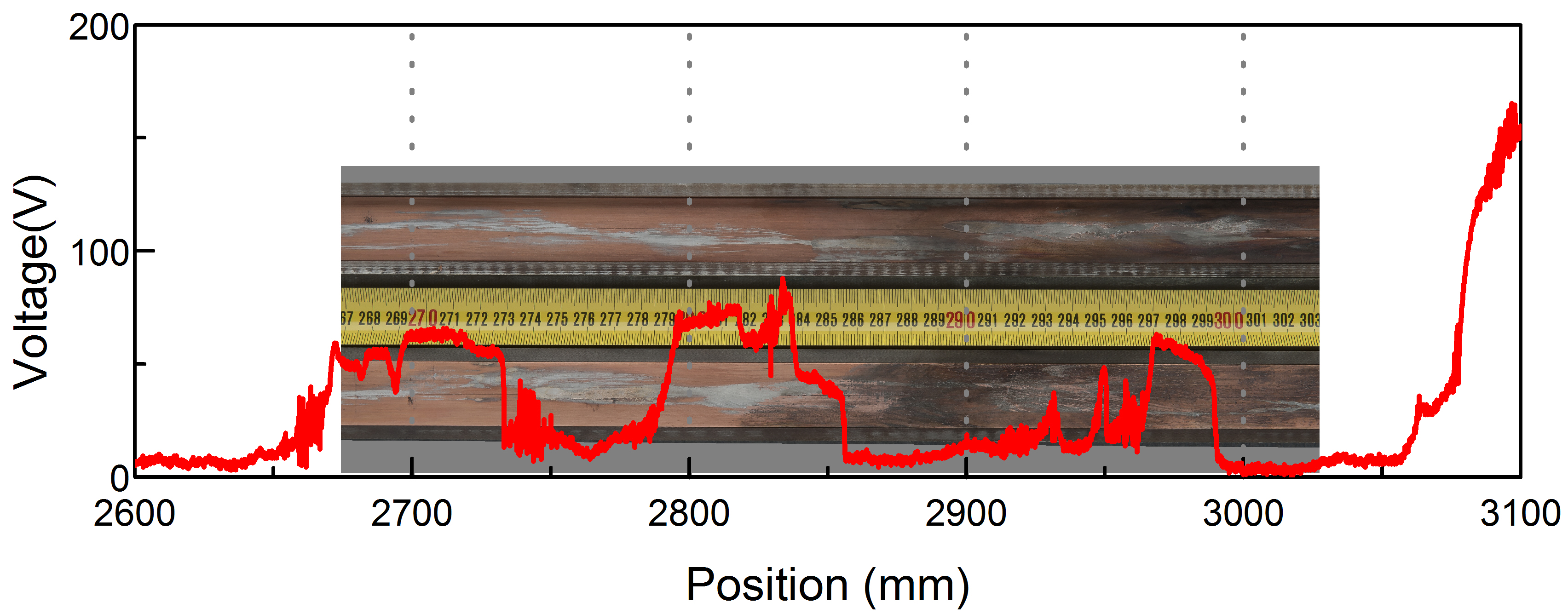}	
    \caption{Extract of the muzzle voltages as function of the z-position with corresponding images of the post shot rail surfaces laid under the muzzle voltage profiles, E = 0.81 MJ, C-shaped projectile}
  \label{fig_contact}
\end{figure}

In the case of the C-shaped projectiles the post shot rail surfaces show a more or less uniform layer of aluminum on the rails, whose thickness increases with increasing primary energy (figure \ref{fig_rails}(a) and (b)).  At positions where peaks presumably caused by current switching between local contact zones arise, this layer is no longer uniform anymore but instead shows distinct traces. Additionally, the color of the layer can get darker in that region (figure \ref{fig_contact}). This fits well to the interpretation that the peaks are caused by current switching between local contact zones.

In the case of the brush projectiles almost no molten copper can be found on the rails. The post shot rail surfaces do only show sliding traces of the brushes before the muzzle voltage peaks occur. At the end of the rails, traces being typical for moving arcs can be seen before the pattern changes due to sliding traces of the brushes back again (see figure \ref{fig_rails}(c) and (d)) \cite{Schneider2003a, Schneider2005}. 

\section{Loss analysis }
During a railgun experiment the armature undergoes two main  sources of losses: The losses due to friction and the losses due to ohmic heating. Both types of losses will be analyzed quantitatively in this section. 
\subsection{Frictional losses}

\begin{figure}
  \centering
    \includegraphics[width=\columnwidth]{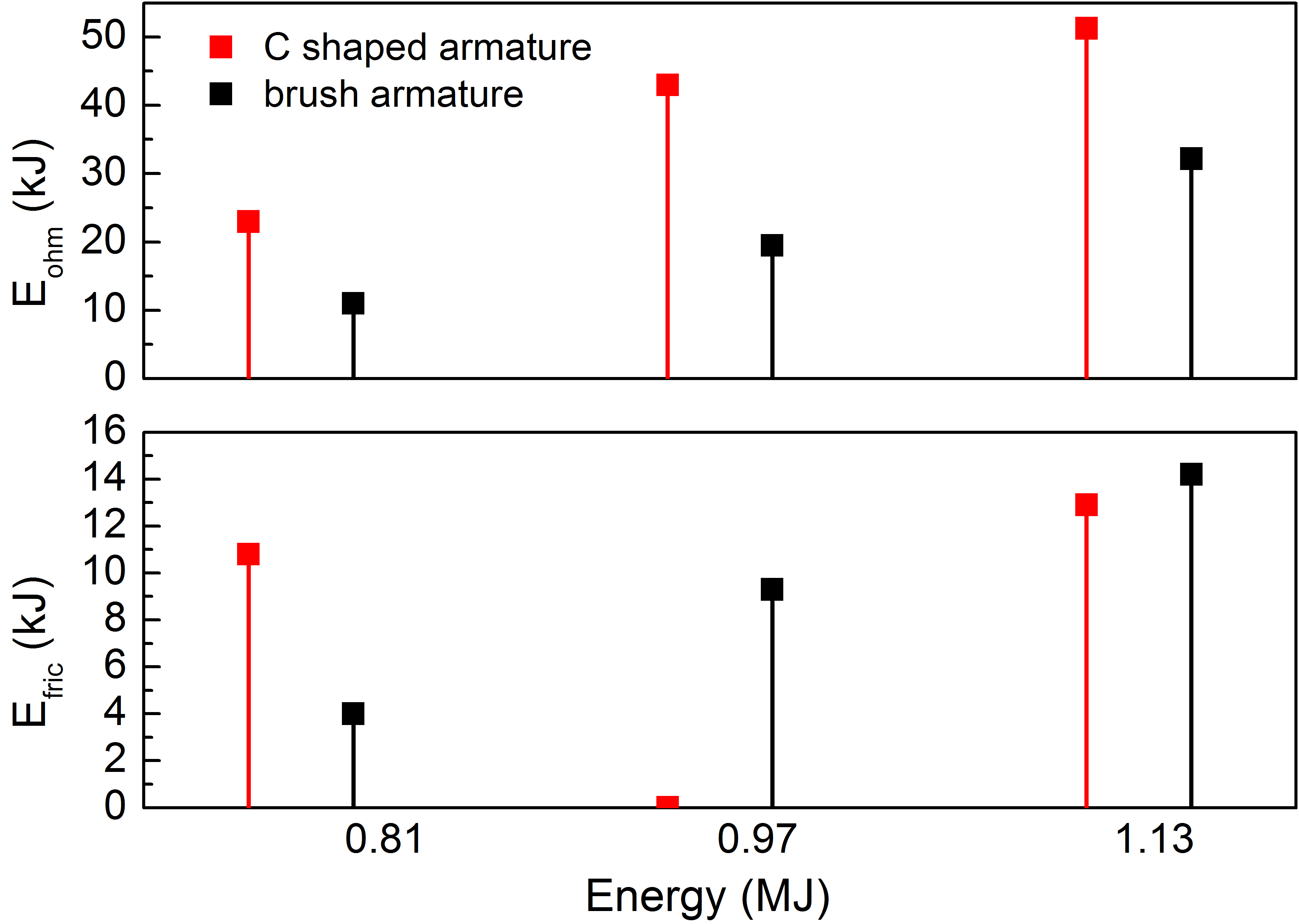}
    \caption{Calculated energy $E_{fric}$ dissipated due to friction (bottom) and 
		Calculated ohmic losses $E_{ohm}$ (top)for the C shaped and brush armature as a function of the applied primary energy}
  \label{fig_e_alle}
\end{figure}

The frictional losses of the two types of armatures can be quantified by the energy $E_{fric}$:
\begin{equation}
\label{eq_efric}
	E_{fric}= E_{kin_{theo}}-E_{kin_{exp}}
\end{equation}
where
\begin{equation}
\label{equ_ekin}
 E_{kin_{theo(exp)}} = \frac{1}{2} m v_{theo(exp)}^2
\label{eq_ekin}
\end{equation}
is the theoretical (experimental) kinetic energy of the projectile with mass $m$ and the corresponding velocity $v$. The values of the experimental velocities are shown in table \ref{tab_velo}. The theoretical velocities are calculated by    :                                                                                                                                                    
\begin{equation}
	v_{theo}= \frac{1}{2m}L'\int_{t_0}^{t_{out}} I^2 dt
	\label{eq_vtheo}
\end{equation}
where $L'$ = 0.45 $\mu$H/m is the inductance gradient along the rails, and $\int_{t_0}^{t_{out}} I^2 dt$  is the action integral, with  $t_0 = 0$ ms  the start time of the shot and $t_{out}$  the exit time of the projectile of the main launcher. In table \ref{tab_velo} the values of the theoretical velocities (calculated with equation \ref{eq_vtheo}) are listed as well.
For both equations, (\ref{eq_efric}) and (\ref{eq_ekin}) no mass reduction is assumed during the experiment.
The results of the calculations are shown as a function of the applied energy in figure \ref{fig_e_alle}. 

For the lowest applied energy (E~=~0.81 MJ), $E_{fric}$ is higher for the C-shaped projectile than for the brush projectile. This might be explained by the fact  that the flexible brushes can adapt better to the unevenness of the rail surfaces than the trailing arm of the C-shaped armature \cite{Wild2015}, although the sliding friction might be lowered  by the molten aluminum on the rails. 

In the case of  an applied energy of E~=~0.97 MJ, the dissipated energy due to friction is smaller for the C-shaped armature than for the brush armature. Furthermore, E$_{fric}$  is smaller than in the previous shot. Due to the higher applied energy, more armature material melts during the shot leading to a significant decrease of the friction at the rail/armature interface, which  is confirmed by the state of the post shot rail surfaces (figure \ref{fig_rails}): the layer of molten aluminum on the rails  seems much thicker than in the case of E~=~0.81 MJ. Furthermore, the very low value of $E_{fric}$ (see figure \ref{fig_e_alle}) can only be explained by a significant mass loss of the armature caused by the melting of the aluminum which might have attributed to the higher obtained velocities as well. 

In the case of the brush armature, $E_{fric}$ is higher than in the previous experiment.  This observation might be explained by an increase of the normal electromagnetic force on the brushes pushing the brushes against the rails \cite{Schneider2003}, leading to a  higher friction coefficient. Additionally,  copper has  a higher   action to melt and  a higher melting point (action to melt = 80490 A$^2$s/mm$^4$, T = 1083 $^{\circ}$C)  than  aluminum (action to melt\,=~25240 A$^2$s/mm$^4$, T = 660 $^{\circ}$C) \cite{Marshall2004a}, no  uniform liquid layer of armature material is formed to lower the friction coefficient  as it is in the case of the C-shaped projectile (see figure \ref{fig_rails}).
  
For the highest applied energy, $E_{fric}$ is again smaller for the C-shaped armature than for the brush armature, but much higher than in the previous shots. During this shot a part of the trailing arms (see figure \ref{fig_rx_rail}(b)) was lost during the launch one can assume that both mass and tribological behavior changed, causing a higher $E_{fric}$. In the case of the brush armature $E_{fric}$ is higher than in the previous experiments, probably due to the same reasons mentioned above.   

\subsection{Ohmic losses}

When electric current flows in a normal conductor, the conductor's temperature will rise because of the resistive
energy loss and/or ohmic losses.
In case of the armatures, this energy loss, $E_{ohm}$, is be described by:
\begin{equation}
E_{ohm} = \int_{t_0}^{t}R_a I^2 dt= \int_{t_0}^{t}U_M I  dt
\label{eq_ohmic_all}
\end{equation}
where $R_a$ is the resistance of the armature, U$_{M}$ the measured muzzle voltage and  $I$  the applied current pulse.  
 With the limits of the integral $t_0=0$ ms and  $t = t_{out}$, the integral of equation (\ref{eq_ohmic_all}) is calculated for all six shots.

The results of these calculations are shown  as a function of the applied energy in figure \ref{fig_e_alle}. For each armature type the ohmic losses increase with increasing applied energy, because of the higher applied current. Moreover, the ohmic  losses $E_{ohm}$ are always greater for the C-shaped armatures than for the brush armatures, which is true for all applied energies. According to paragraph \ref{sub_muzzle} and equations (\ref{equ_ohm}) and (\ref{equ_r}) this can be explained  by the longer current path in the C-shaped armature and by the higher electrical resistivity of aluminum.

\subsection{Total losses}

\begin{figure}
  \centering
    \includegraphics[width = \columnwidth]{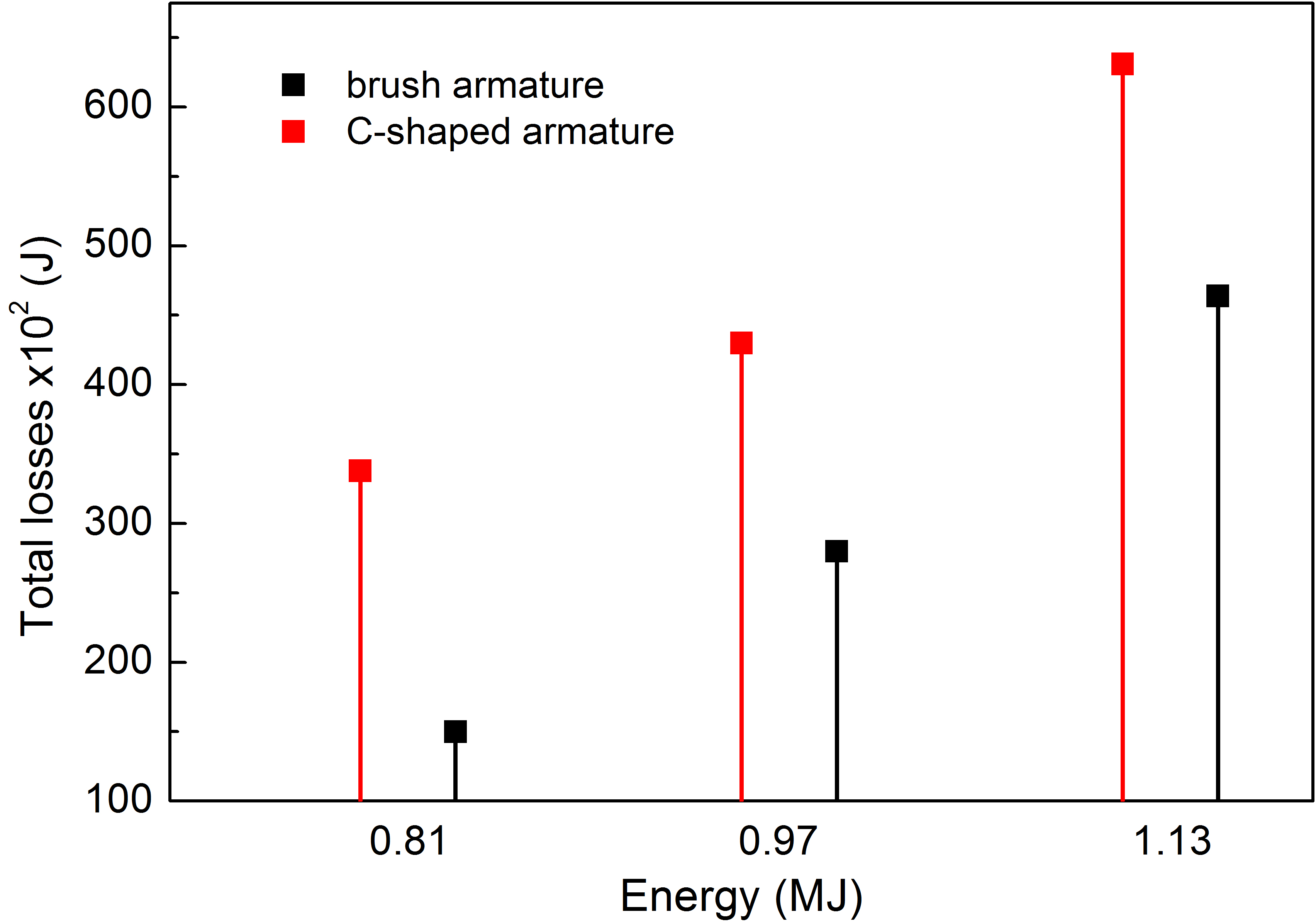}
    \caption{Sum of the ohmic losses $E_{ohm}$ and losses due to friction $E_{fric}$ for the C shaped and brush projectile as a function of the applied primary energy}
  \label{fig_e_all}
\end{figure}
 In order to compare the overall losses of both armatures the total energy loss, $E_{loss}$, is calculated:
\begin{equation}
E_{loss} = E_{ohm} + E_{fric}
\label{eq_total_loss}
\end{equation}
In figure \ref{fig_e_all} the results of equation (\ref{eq_total_loss}) are shown for all shots as a function of the applied energy:  for both armature technologies the total losses increase  with the primary energy, because the ohmic losses and frictional losses, as already mentioned above, increase  with the applied energy as well.
But the total losses of the brush armatures are about 2/3 of the total losses of the C-shaped technology at maximum, i.e. the smaller ohmic losses of the brushes cannot be compensated by the lower frictional losses of the C-shaped armature. Therefore the total losses of the brush armatures are always smaller than the total losses of the C-shaped armatures, because of the higher ohmic losses of the latter one.
 
\section{Efficiency}
\begin{figure}
  \centering
    \includegraphics[width=\columnwidth]{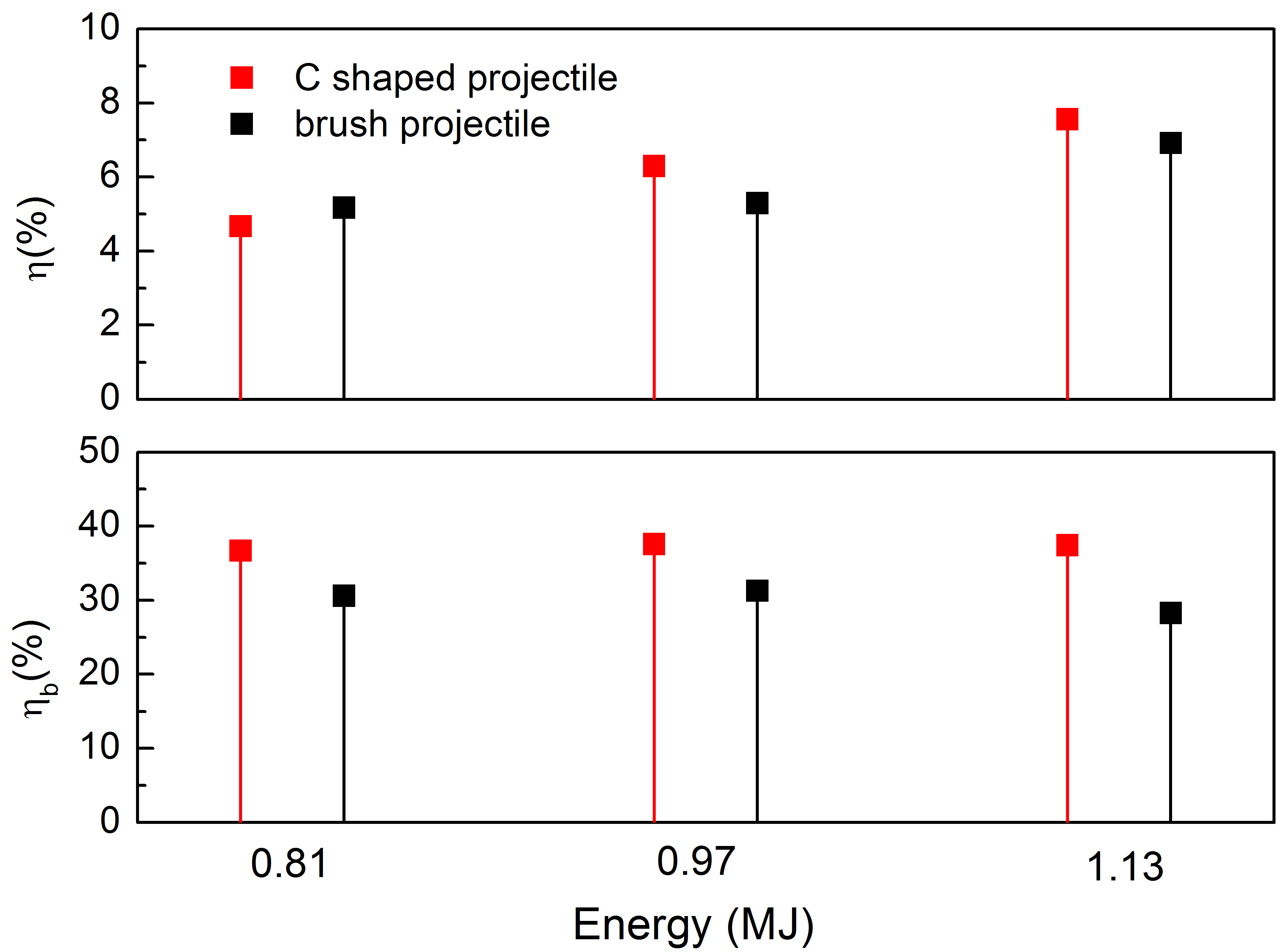}
    \caption{Efficiencies $\eta$ (top) and $\eta_b$ (bottom) of the C-shaped and brush projectile as a function of the applied primary energy}
  \label{fig_eff}
	 \end{figure}

One important parameter to quantify the performance of the two types of armature technologies is the efficiency, i.e. how the applied electrical energy is converted into kinetic energy.  The system efficiency $\eta$ can be defined as:
\begin{equation}
\eta= \frac{E_{kin_{exp}}}{E_{total}}
\end{equation}
where E$_{total}$ is the initially stored energy in the capacitors and E$_{kin_{exp}}$ is the experimental kinetic energy  defined in equation (\ref{eq_ekin}). With the values of $v_{exp}$ listed in table \ref{tab_velo}, $\eta$ is calculated for each experiment and the results are shown in figure \ref{fig_eff}.  
The efficiency increases with velocity for both types of projectiles corresponding to a well-known characteristic of a DC-driven railgun \cite{Marshall2004a}. For the lowest applied energy, $\eta$ is lower for the C-shaped projectile than for the brush projectile (about 10\%), because of the lower velocity of the C-shaped projectile. For all other applied energies it is the other way around, because in these experiments the velocities of the C-shaped projectiles where greater than those for the brush projectiles (about 16\%). 
	
	Another way to define the efficiency  is to replace the energy stored in the capacitor by the energy, $E_{breech}$, delivered  to the current
injection points, not taking into account any auxiliary systems. We therefore write for this efficiency called $\eta_b$: 
\begin{equation}	
	\eta_b = \frac{E_{kin_{exp}}}{E_{breech}}
	\label{equ_etab}
\end{equation}
with the quantity $E_{breech}$ 
\begin{equation}
	E_{breech}=\int_{t_0}^{t_{out}} U_{breech}I dt
\end{equation}
$U_{breech}$ is the breech voltage defined in  \cite{Marshall2004a}.

 %and defined as:
%\begin{equation}
	%U_{breech}=\int_{t_0}^{t_{out}} U_{breech}I dt
%\end{equation}

The values of $E_{breech}$ are listed in table \ref{tab_ebreech}. The efficiency $\eta_b$ is calculated for all performed shots and the results of these calculations are shown in figure \ref{fig_eff}. Similar to the quantity $\eta$, the efficiency $\eta_b$ increases  with velocity for both types of  projectiles \cite{Marshall2004a}. Furthermore, $\eta_b$ is always greater for the C-shaped projectile than for the brush projectile (up to 25 \%). According to \cite{Marshall2004a} the breech voltage, and consequently $E_{breech}$  as well, depends amongst other things on the electrical resistivity of the rails. Since we use copper rails  in the case of the C-shaped projectile and aluminum rails in case of the brush projectile,  $E_{breech}$ is always lower (see table \ref{tab_ebreech}) for the C-shaped projectiles than for the brush projectiles (up to 25 \%). In combination with the experimental velocities, higher values of $\eta_b$ for the C-shaped projectile than for the brush projectile are thus obtained (see equation (\ref{equ_etab})). Therefore we can conclude that the projectiles equipped with  C-shaped armature converted more efficiently electrical in kinetic energy than the projectiles equipped with brush armatures. 

%The last efficiency calculated is the armature efficiency. 
%Due to the low specific mass of the aluminum used for the C-shaped projectile,  

  \begin{table}
		\centering
        \caption{$E_{breech}$, energy delivered to the launcher }
        \label{tab_ebreech}
        \begin{tabular} {| c | c c |}
            \cline{1-3}
						 \cline{1-3}
           & \multicolumn{2}{|c|}{\textbf{$E_{breech}$ (kJ)}}\\
            \cline{1-3}
						 %\cline{1-3}
              \textbf{Energy (MJ)} & C-shape & brush \\
						\hline
          	0.81 &	10.3 & 13.7 \\
						
			\hline%
				0.97 & 16.3 & 16.9 \\
			\hline%
				1.13 & 22.8 & 27.6 \\
            \hline
						 \cline{1-3}
        \end{tabular}
    \end{table}	
%with $E_{total}$ the energy initially stored in the capacitors and $E_{kin_{exp}}$ the experimental kinetic energy (see equation \ref{eq_ekin}). Both efficiencies  to increase with velocity  for both types of armatures corresponding to a well-known characteristic of a DC-driven railgun \cite{Marshall2004a}. The difference  between $\eta$  and $\eta_b$ in absolute values points to the losses in the electric circuit and to the energy dissipated after the exit of the projectile. 
%By  comparing  the performance of both technologies, they are in the same order of magnitude. However, the comparison between the armature types is more appropriate if corrected with respect to the mass of the armature. The C-armature used here has a mass of approximately 30 g (i.e. the C shaped part of the projectile), whereas the brush armatures have a total mass of about 40 g (i.e. the mass of all six copper brushes).  Consequently, C shaped armatures mad of aluminum convert more efficiently (about $\approx$25\%)  electrical energy into kinetic energy than multiple copper brush armatures. This superiority can  mainly be attributed to the lower density of aluminum ($\rho_{Al} = 2.7~g/cm^3$ and/or $\rho_{Cu} = 9.0~g/cm^3$). 

\section{Summary and conclusions}

In this paper we performed a series of experiments with projectiles equipped either with C-shaped  armatures (made of aluminum) or with brush armatures (made of copper). These projectiles were constructed in such a way, to be as similar as possible in terms of payload and armature mass.  But because of the material parameter $\rho /\sqrt{g_1}$  the mass of copper armatures is always   bit   than the mass of aluminum for a given action integral. The total mass of both types of projectiles was always 80~g. With the railgun RAFIRA we performed single shots with increasing primary energy (E= 0.81 MJ, 0.97 to 1.13 MJ). The experimental results obtained during the shot were analyzed qualitatively and quantitatively.

Both types of projectiles showed an overall good  electrical contact behavior. In the case of the C-shaped projectile a contact transition was observed once at the end of the shot (at E~=~1.13 MJ).  For both types of projectiles we observed  narrow peaks peaks with amplitudes up to 80~V  at the end of the shots, indicating a temporary contact loss of the armature. These peaks might be attributed to current switching process either between brushes or local contact zones in case of the C-shaped armature.

 In the case of the C-shaped projectiles the post shot rail surfaces showed layers of molten aluminum. In the case of the brush projectiles typical sliding traces of the multiple brushes could be observed on the post shot rail surfaces. For both technologies distinctive traces due to a temporary contact losses could be found on the rail surfaces.  

 For both  types of projectiles we obtained velocities in the range between  950~m/s and 1450 ~m/s. Except for the lowest applied energy the velocities reached with the C-shaped projectile were always greater than those obtained with brush projectiles ($\Delta v_{max} = 70~m/s$).  

The losses due to friction were always, except for the lowest applied energy, lower for the C-shaped armature than for the brush armature. 
This might be explained by the lower melting point and action to melt of aluminum causing a more or less uniform layer of molten aluminum on the rail lowering the frictional losses. In addition, the caused mass loss of the C-shaped armature due to the melting of the aluminum might have attributed to the higher obtained velocities as well.  

In the case of the ohmic losses, the brush armatures showed lower ohmic losses than the C-shaped armatures, because of the lower electrical resistivity of copper and the shorter current path through the brush armature.  
 
Nevertheless, the total losses (sum of ohmic and frictional losses) of the brush armatures amounted to 2/3 of the C-shaped armatures at maximum. Therefore, the smaller ohmic losses of the brushes cannot be compensated by the lower frictional losses of the C-shaped armatures. 

Finally, we determined the efficiencies $\eta$ and $\eta_b$ by comparing the kinetic energy to $E_{total}$ (the energy initially stored in the capicators) and $E_{breech}$ (the energy delivered to the current injection points), respectively. Both efficiencies increased with  increasing primary energy, because of the increasing muzzle velocities. But  $\eta$ and $\eta_b$ are always greater for the projectile with C-shaped armatures  than for the projectiles with brush armatures, because of the lower electrical resistivity of the used copper rails used for the projectiles with C-shaped armatures. Consequently, the projectiles with C-shaped armature converted better the electrical energy into kinetic energy than the projectiles with brush armatures.

\bibliographystyle{IEEEtran}
\bibliography{IEEEabrv,eml}

\end{document}